\documentclass[aps,prl,reprint,floatfix]{revtex4-2}

\usepackage[utf8]{inputenc}
\usepackage[T1]{fontenc}

\usepackage{amsmath}
\usepackage{amssymb}
\usepackage{amsfonts}
\usepackage{graphicx}
\usepackage{siunitx}
\usepackage{xcolor}
\usepackage{physics} 

\usepackage[colorlinks=true,citecolor=blue,urlcolor=red]{hyperref}
\usepackage[normalem]{ulem}

\newcommand{\W}{\mathcal{W}}
\newcommand{\Q}{\mathcal{Q}}

\begin{document}
\title{Information and thermodynamics: fast and precise approach to Landauer's bound in an underdamped micro-mechanical oscillator}
\author{Salamb\^{o} Dago}
\author{Jorge Pereda}
\author{Nicolas Barros}
\author{Sergio Ciliberto}
\affiliation{Univ Lyon, ENS de Lyon, Univ Claude Bernard Lyon 1, CNRS, Laboratoire de Physique, F-69342 Lyon, France}
\author{ Ludovic Bellon}
\email{ludovic.bellon@ens-lyon.fr}
\affiliation{Univ Lyon, ENS de Lyon, Univ Claude Bernard Lyon 1, CNRS, Laboratoire de Physique, F-69342 Lyon, France}

\begin{abstract}
The Landauer principle states that at least $k_B T \ln 2$ of energy is required to erase a 1-bit memory, with $k_B T$ the thermal energy of the system. We study the effects of inertia on this bound using as one-bit memory an underdamped micro-mechanical oscillator confined in a double-well potential created by a feedback loop. The potential barrier is precisely tunable in the few $k_B T$ range. We measure, within the stochastic thermodynamic framework, the work and the heat of the erasure protocol. We demonstrate experimentally and theoretically that, in this underdamped system, the Landauer bound is reached with a $\SI{1}{\%}$ uncertainty, with protocols as short as $\SI{100}{ms}$.
\end{abstract}

\maketitle

The thermodynamic energy cost of information processing is a widely studied subject both for its fundamental aspects and for its potential applications~\cite{Rex, Parrondo_sagawa, CiliLutz_PT, Orlov_book, Gammaitoni_2016, Lutz_2021,Toyabe2010,Parrondo2014,Pekola2014}. This energy cost has a lower bound, fixed by Landauer’s principle~\cite{Landauer_1961}: at least $k_BT \ln2$ of work is required to erase one bit of information from a memory at temperature $T$, with $k_B$ the Boltzmann constant. This is a tiny amount of energy, only $\sim \SI{3 e-21}{J}$ at room temperature ($\SI{300}{K}$), but it is a general lower bound, independent of the specific type of memory used, and it is related to the generalized Jarzynski equality~\cite{BerutEPL2013}. The Landauer bound (LB) has been measured in several classical experiments, using optical tweezers~\cite{Berut2012,Berut2015}, an electrical circuit~\cite{orl12}, a feedback trap~\cite{Bech2014,Gavrilov_EPL_2016,Finite_time_2020} and nanomagnets~\cite{Hong_nano_2016,mar16} as well as in quantum experiments with a trapped ultracold ion~\cite{Yan_2018} and a molecular nanomagnet~\cite{gau17}. The LB can be reached asymptotically in quasi-static erasure protocols whose duration is much longer than the relaxation time of the above mentioned systems used as one-bit memories. In practice, when the erasure is performed in a short time, the energy needed for such a process can be minimized using optimal protocols, which have been computed~\cite{Aurell_2012,Diana_2013,Schmiedl_2007,Boyd_PRX,Boyd_arx,Muratore-Ginanneschi-2017} and used for overdamped systems~\cite{Finite_time_2020}. Another strategy to approach the asymptotic LB faster is of course to reduce the relaxation time. However, for very fast protocols, one may wonder whether inertial (inductive) terms in mechanical (electronic) systems play a role in their reliability and energy cost. 

\begin{figure}
	\centering
	\includegraphics[width=80mm]{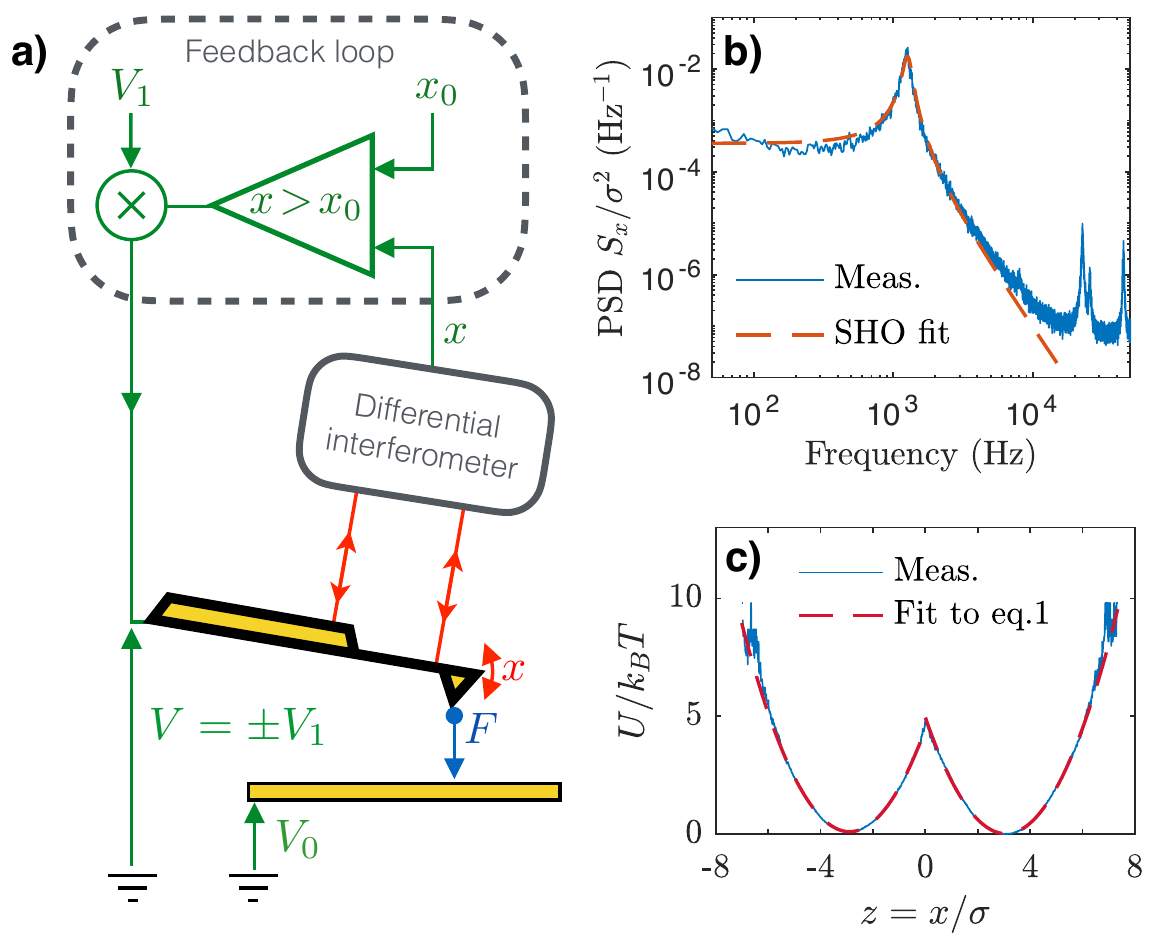}
	\caption{(a) Schematic diagram of the experiment. The conductive cantilever is sketched in yellow. Its deflection $x$ is measured with a differential interferometer~\cite{Paolino2013}, through two laser beams focused respectively on the cantilever and on its base. The cantilever at voltage $V=\pm V_1$ is facing an electrode at $V_0$. The voltage difference $V-V_0$ between them creates an attractive electrostatic force $F\propto(V-V_0)^2$. The dashed box encloses the feedback controller made of a comparator and a multiplier, which create the double-well potential. (b) Measured Power Spectrum Density (PSD) of the cantilever deflection thermal noise with no feedback ($V_1=0$, blue), and best fit by the theoretical thermal noise spectrum of a Simple Harmonic Oscillator (SHO, dashed red). Up to $\SI{10}{kHz}$, the cantilever behaves like a resonator at $f_0=\SI{1270}{Hz}$, with a quality factor $Q=10$. We infer from this measurement the variance $\sigma^2 = \langle x^2 \rangle = k_B T / k$, used to normalize all measured quantities. (c) Measured double-well potential energy (blue) obtained with the feedback on, with $x_0=0$ and $V_1$ adjusted to have a $5\,k_BT$ barrier height. The potential is inferred from the measured Probability Density Function (PDF) of $x$ during a $\SI{10}{s}$ acquisition and the Boltzmann distribution. The fit using Eq.~\eqref{eq_U(z,z0,z1)} is excellent (dashed red).}
	\label{schema_bloc}
\end{figure}

\begin{figure*}
 \centering
 \includegraphics[width=0.92\textwidth]{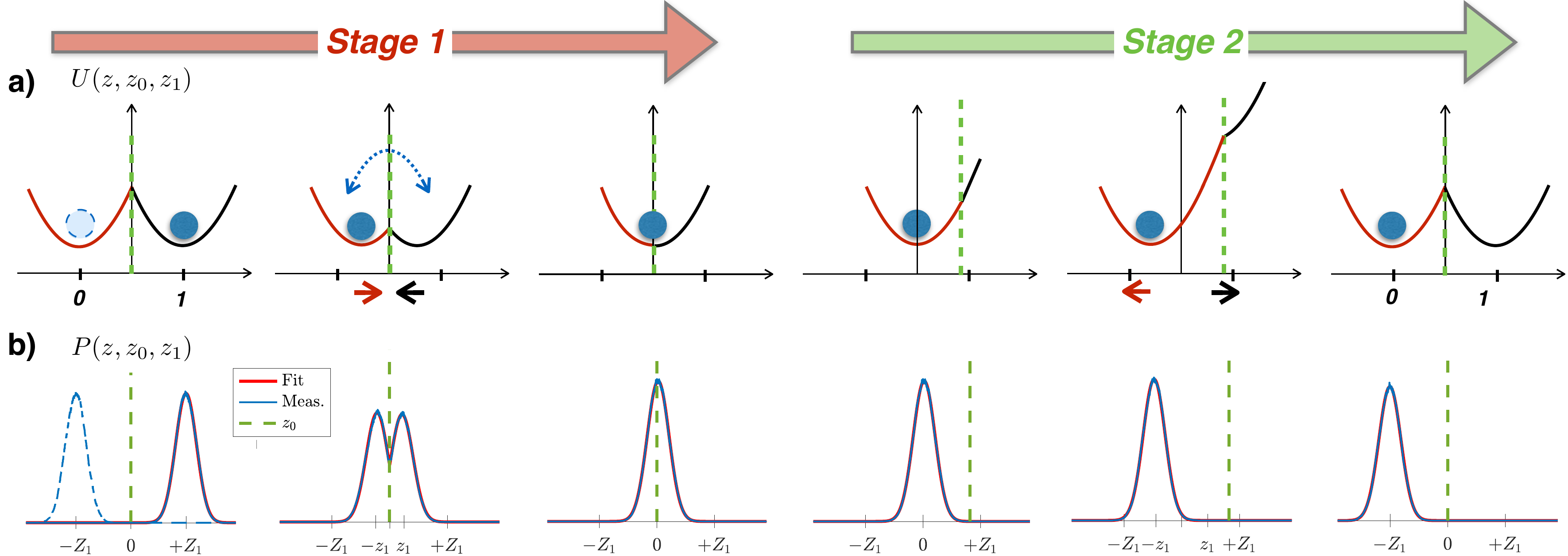}
\caption{Description of the erasure protocol. (a) Evolution of the potential $U(z,z_0,z_1)$. (b) Experimental static PDF $P(z,z_0,z_1)$ (blue), 
	best fit to the Boltzmann distribution (red), and position of the threshold $z_0$ (dashed green).}
 \label{schema_protocol}
\end{figure*}

The goal of this letter is to analyze this problem and to provide a new experimental and theoretical exploration of Landauer's principle on an underdamped micro-mechanical oscillator confined in a very specific double-well potential. Erasure procedures in underdamped systems have never been studied before, and this new application field opens significant possibilities, since oscillators are fundamental building blocks to many systems. Moreover, it is interesting to verify the LB for a weak coupling to the thermostat. Both the relaxation time and the coupling to the bath of our system are orders of magnitude smaller than those of the overdamped systems of previous demonstrations. We approach the LB in $\SI{100}{ms}$, compared to the $\SI{30}{s}$~\cite{Berut2015} previously needed, and thus accumulate much more statistics. Notice that the asymptotic time in our experiment is among the fastest erasure times of the systems mentioned above. Furthermore, our choice of confining potential allows both a precise experimental control and an analytical computation of the work and heat on a trajectory. These predictions can be compared quantitatively with the experimental results because they include only quantities that can be measured with high precision in our experimental set-up, schematically described in Fig.~\ref{schema_bloc}(a).

The underdamped oscillator is a conductive cantilever~\footnote{Doped silicon cantilever OCTO1000S from Micromotive Mikrotechnik, nominal stiffness $\SI{3e-3}{N/m}$.} which is weakly damped by the surrounding air. Its deflection $x$ is measured with very high accuracy and signal-to-noise ratio by a differential interferometer~\cite{Paolino2013}. The Power Spectrum Density (PSD) of the thermal fluctuations of $x$ is plotted in Fig.~\ref{schema_bloc}(b). The fit of this PSD with the thermal noise spectrum of a Simple Harmonic Oscillator (SHO) gives its resonance frequency $f_0=\omega_0/2\pi=\SI{1270}{Hz}$ and quality factor $Q=m\omega_0/\gamma=10$, where $m$, $k=m \omega_0^2$ and $\gamma$ are respectively: the mass, stiffness and damping coefficient of the SHO. From the PSD we compute the variance $\sigma^2 = \langle x^2 \rangle = k_B T / k \sim \SI{1}{nm^2}$, which is used to normalize all measured quantities: deflections are from here on expressed as $z=x/\sigma$, and energies are in units of $k_B T$. The SHO potential energy is for example $\frac{1}{2} k x^2 / k_B T= \frac{1}{2} z^2$, and its kinetic energy $E_k=\frac{1}{2} m \dot{x}^2 / k_B T= \frac{1}{2} \dot{z}^2 / \omega_0^2$. 

In order to use the cantilever as a one-bit memory, we need to confine its motion in an energy potential consisting of two wells separated by a barrier, whose shape can be tuned at will. This potential $U$ is created by a feedback loop, which compares the cantilever deflection $z$ to an adjustable threshold $z_0$. After having multiplied the output of the comparator by an adjustable voltage $V_1$, the result is a feedback signal $V$ which is $+V_1$ if $z>z_0$ and $-V_1$ if $z<z_0$. The voltage $V$ is applied to the cantilever which is at a distance $d$ from an electrode kept at a voltage $V_0$. The cantilever-electrode voltage difference $V_0\pm V_1$ creates an electrostatic attractive force $F=\frac{1}{2}\partial_dC(d)(V_0 \pm V_1)^2$~\cite{Butt-2005}, where $C(d)$ is the cantilever-electrode capacitance. Since $d\gg \sigma$, $\partial_dC(d)$ can be assumed constant. We apply $V_0\sim \SI{100}{V}$ and $V_1\ll V_0$ so that in a good approximation, $F\propto \pm V_1$ up to a static term. This feedback loop results in the application of an external force whose sign depends on whether the cantilever is above or below the threshold $z_0$. The reaction time of the feedback loop is very fast (around $\num{e-3}f_0^{-1}$), thus the switching transient is transparent thanks to the inertia of the cantilever. As a consequence, the oscillator evolves in a virtual static double-well potential, whose features are controlled by the two parameters $z_0$ and $V_1$. Specifically, the barrier position is set by $z_0$ and its height is controlled by $V_1$ which sets the wells centers $\pm z_1 = \pm V_1 \partial_dC(d)V_0/ (k \sigma)$. The potential energy constructed by this feedback is:
\begin{align}
U(z,z_0,z_1)=\frac{1}{2} \big(z-S(z-z_0)z_1\big)^2,
\label{eq_U(z,z0,z1)}
\end{align}
where $S(z)$ is the sign function: $S(z)=-1$ if $z<0$ and $S(z)=1$ if $z>0$. We carefully checked that the feedback loop creates a proper virtual potential without perturbing the response of the system, including having no effect on the equilibrium velocity distribution \cite{suppmat}. The potential in Eq.~\eqref{eq_U(z,z0,z1)} can be experimentally measured from the Probability Distribution Function (PDF) $P(z$) and the Boltzmann distribution $P(z)\propto e^{-U(z)}$. Figure \ref{schema_bloc}(c) presents an example of an experimental symmetric double-well potential generated by the feedback loop, tuned to have a barrier of $\frac{1}{2}z_1^2 = 5$ ($z_0=0$, $z_1=\sqrt{10}$). The dashed red line is the best fit with Eq.~\eqref{eq_U(z,z0,z1)}, demonstrating that the feedback-generated potential behaves as an equilibrium one.

The erasure protocol corresponds to the potential evolution described in Fig.~\ref{schema_protocol}, with the associated experimental static position distribution $P(z)$. Initially, the system is at equilibrium either in the state 0 (left-hand well centered in $-Z_1$) or in the state 1 (right-hand well centered in $Z_1$) with a probability $p_i=\frac{1}{2}$. The process must result in the final state 0 with probability $p_f=1$, to perform a full erasure process. To do so, during the first stage we drive the wells closer until they merge into one single harmonic well centered in $0$. After a short equilibration time, the barrier position (dashed green line) is pushed away to prevent the cantilever from visiting the right hand well (black line). It thus remains in the left hand well (red line), which is driven back to its initial position in $-Z_1$. The so-called stage 2 ends when the threshold is brought back in $0$, to reach the final state 0 in the bi-stable potential, regardless of the initial state.

\begin{figure}[h]
 \includegraphics[width=85mm]{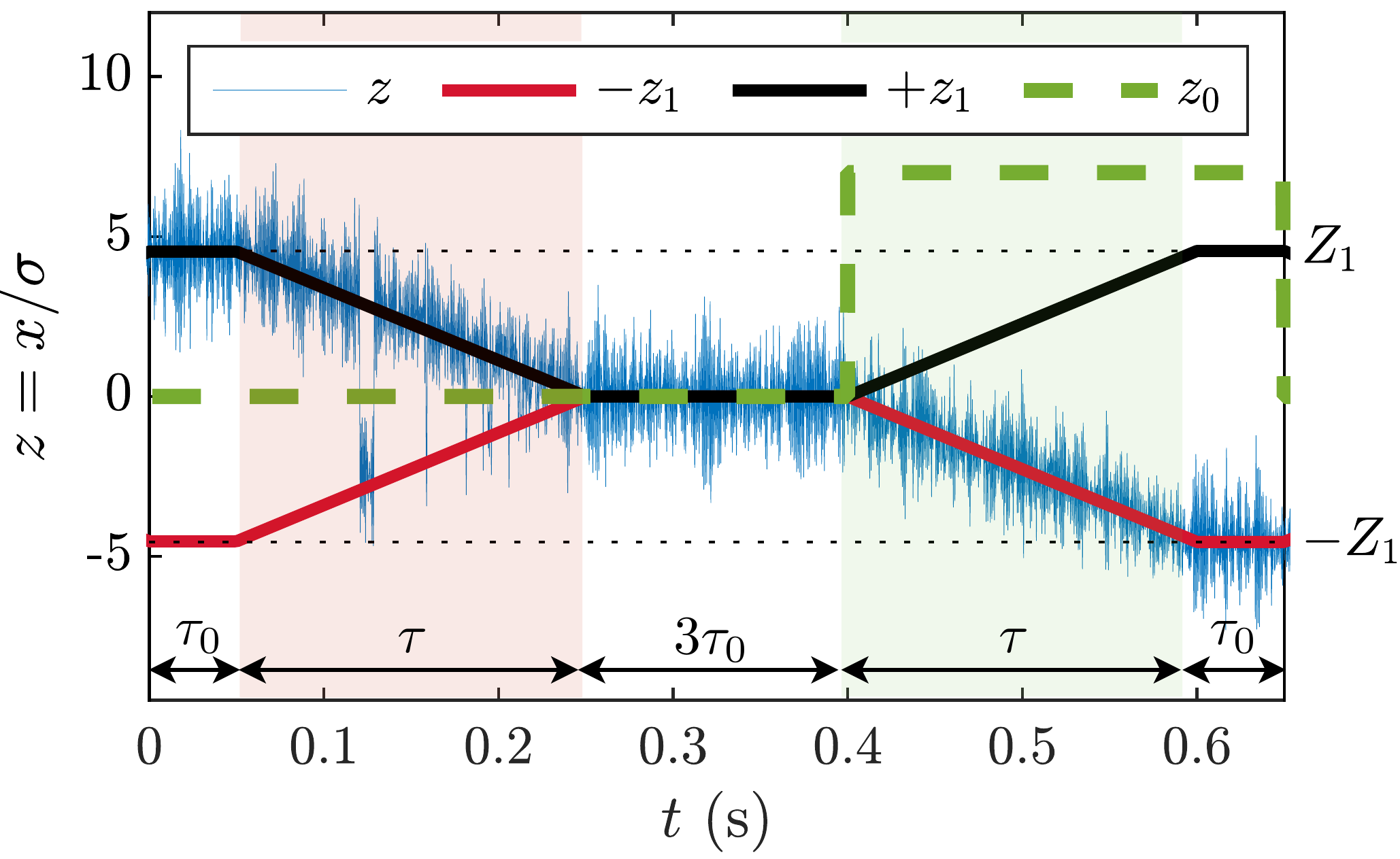}
 \caption{Time recording of the cantilever deflection $z$ on a single trajectory (blue, starting in state 1 in this example), superposed with the two wells' centers $+z_1$ (black) and $-z_1$ (red), and the threshold $z_0$ (dashed green). Stage 1 (red background) et 2 (green background) both last $\tau=\SI{200}{ms}$. $\tau$ is very long compared to the natural relaxation time $\tau_r$ of the cantilever: $f_0\tau \sim 250 \gg f_0\tau_r = Q/\pi \sim 3$. The equilibrium periods around stages 1 and 2 are chosen freely as long as they allow the cantilever to relax, in this example $\tau_0=\SI{50}{ms}\gg\tau_r$.} \label{cycle_lent}
\end{figure}

To implement the erasure procedure, the position of the center of the wells $\pm z_1$ (black and red lines), and the barrier position $z_0$ (green dashed line) are driven accordingly to the profiles in Fig.~\ref{cycle_lent}. Initial equilibrium is ensured by a steady potential for a duration of $\tau_0$, chosen longer than the natural relaxation time of the system $\tau_{r} = 2 Q / \omega_0 = \SI{2.5}{ms}$. For clarity purposes, $\tau_0=\SI{50}{ms}$ in Fig.~\ref{cycle_lent}. During stage 1 (red background) the center of the wells moves linearly from the initial state $z_1(0)=Z_1=4.55$ to $z_1(\tau)=0$ in a time $\tau=\SI{200}{ms}$. $Z_1$ sets the height of the barrier in the initial state $\frac{1}{2}Z_1^2 \gtrsim 10$, chosen high enough to secure the initial and final states. When the two wells are close enough, the cantilever starts switching from one well to the other: the information is progressively lost. The cantilever is then allowed to relax at equilibrium in this single well for $3 \tau_0$. During stage 2 (green background), the threshold $z_0$ is set at a large value (of order $Z_1$ or higher): the cantilever can no longer switch to the right hand side well. The cantilever is then brought back to the state 0, as $z_1$ follows a linear ramp from $0$ to $Z_1$, in the same time $\tau$. The protocol ends at $\tau_f = 2 \tau + 5 \tau_0 \gtrsim 2 \tau + 15 \tau_r$~\footnote{To set $\tau_0$, we use the rule of thumb that a perturb system returns to equilibrium after $3\tau_r$, with $\tau_r$ the characteristic relaxation time in the exponential decay.}, after letting the system relax during $\tau_0$ and bringing back $z_0$ to 0.

The data plotted in Fig.~\ref{cycle_lent} contains all we need to compute the stochastic work and heat along a single trajectory. Indeed, if we apply to the underdamped regime the generic computations of these stochastic quantities~\cite{sek10,Sekimoto,Seifert_2012,Jarzynski_2011,Ciliberto_PRX}, we obtain the following expressions~\cite{suppmat}:
\begin{align}
\W&=\int_0^{\tau_f} \sum_{i=0,1}\frac{\partial U}{\partial z_i} \dot z_i \dd t = \int_0^{\tau_f} \! \big(S(z\!-\!z_0)z-z_1\big) \dot z_1 \dd t \label{WandQ1}\\
\begin{split}
\Q&=- \int_0^{\tau_f} \frac{\partial U}{\partial z} \dot z \dd t - \Big[ E_k \Big]_0^{\tau_f} \\
&= \int_0^{\tau_f} \big(S(z-z_0) z_1-z \big) \dot z \dd t - \frac{1}{2\omega_0^2} \Big[ \dot{z}^2 \Big]_0^{\tau_f}. \label{WandQ}
\end{split}
\end{align}
From the above definitions, we deduce the energy balance of the system: 
\begin{align}
\Delta U+\Delta E_k=\W-\Q.
\label{energy}
\end{align}
Since at $\tau_f$ the system has relaxed to equilibrium, and the initial and the final states of the protocol are the same, $\langle \Delta U \rangle = \langle \Delta E_k \rangle = 0$, thus in average $\langle \W \rangle = \langle \Q \rangle$~\cite{note}.

\begin{figure}
 \includegraphics[width=80mm]{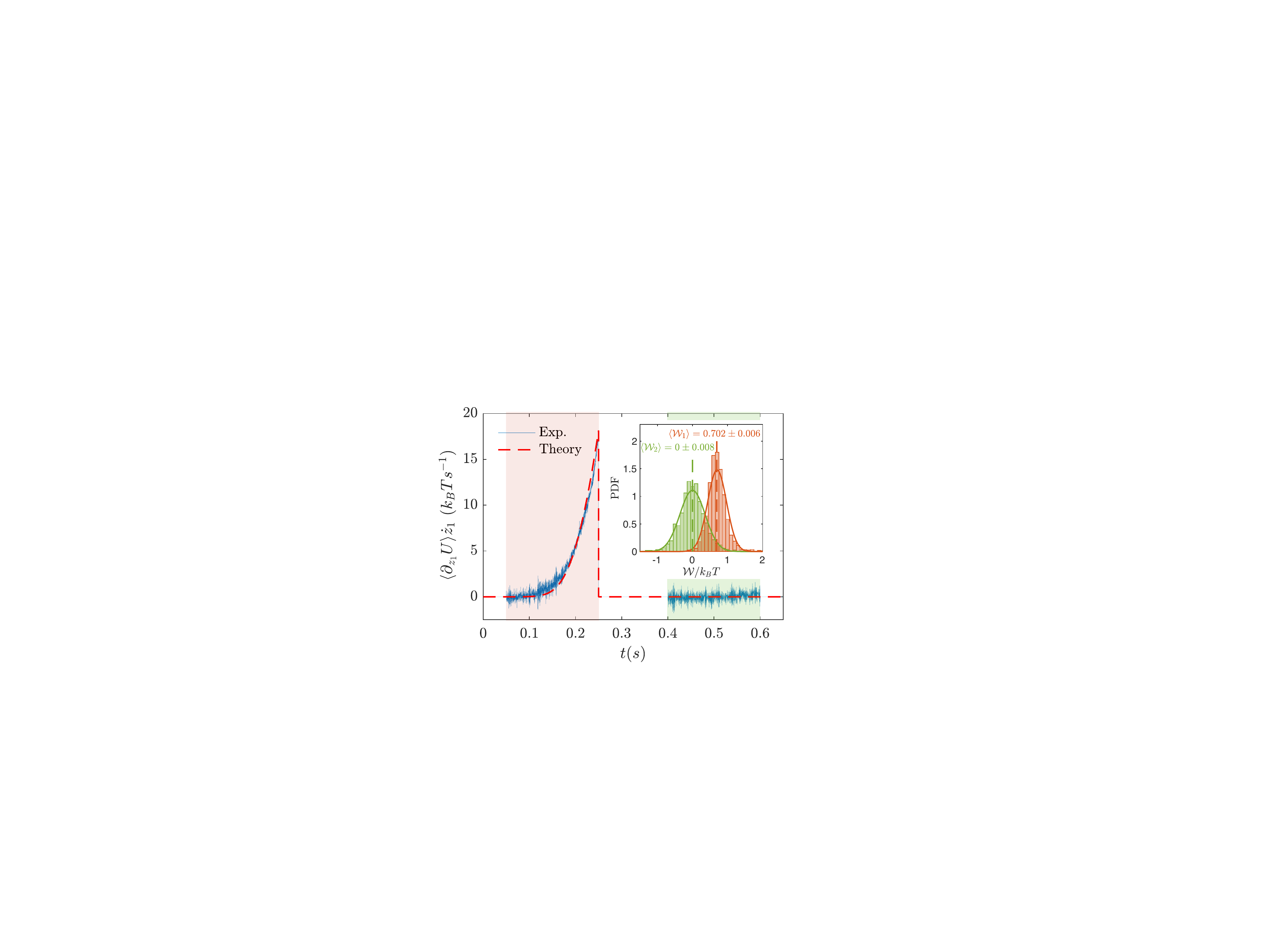}
 \caption{Time evolution of the mean power over 2000 trajectories following the slow protocol of Fig.~\ref{cycle_lent}. The work takes off when the cantilever starts switching between the wells, during stage 1 (red background). Since the process is quasi-static, stage 2 (green background) doesn't contribute on average. The red dashed line is the analytical prediction in the quasi-static regime~\cite{suppmat}. (Inset) Work distributions for the 2000 trajectories during stage 1 ($\W_1$, red) and 2 ($\W_2$, green) fitted by gaussians (plain lines). As theoretically predicted for a slow erasure, stage 1 reaches the LB of $0.693$ (dashed red) while stage 2 requires negligible work (dashed green). PDF of $\Q$ are available in the Suppl. Mat.~\cite{suppmat}}
 \label{dW_lent}
\end{figure}

In order to reach the LB, we should proceed in a quasi-static fashion. From Eqs.~\eqref{WandQ1} and \eqref{WandQ}, using the Boltzmann equilibrium distribution and the potential energy of Eq.~\eqref{eq_U(z,z0,z1)}, we demonstrate that in the quasi-static regime stage 1 requires $\langle \W \rangle = \langle \Q \rangle = \ln 2$, and that stage 2 doesn't contribute \cite{suppmat}. Actually, in a quasi-static evolution, any intermediate state can be theoretically described with our specific choice of $U(z,z_0,z_1)$, and the mean power computed all along the cycle. Long protocols satisfying $\tau\gg\tau_{r}$ can be considered as quasi-static erasures, hence we present in Fig.~\ref{dW_lent} the results obtained from 2000 trajectories, equivalent to the one of Fig.~\ref{cycle_lent}. It should be emphasized that the protocol is perfectly efficient from an information processing point of view: all 2000 trajectories ended in the prescribed well, regardless of the initial condition.

As expected for a slow translation of a single well, stage 2 requires nearly no power: $\langle \W_2 \rangle = 0 \pm 0.008$. During stage 1 however, the power increases when the cantilever starts switching between the wells. Of course, the equilibrium stage never contributes to the erasure cost. Summing the dissipated power along the process gives: $\langle \W_1 \rangle = 0.702 \pm 0.006$ and $\langle \W_2 \rangle = 0 \pm 0.008$. Furthermore, we also compute the mean dissipated heat during the whole procedure: $\langle \Q \rangle=0.68 \pm 0.03$. These results are in perfect agreement with the Landauer principle: in a quasi-static process, the mean work or heat required to erase 1 bit of information is $\ln 2=0.693$. 

\begin{figure}[t]
 \includegraphics{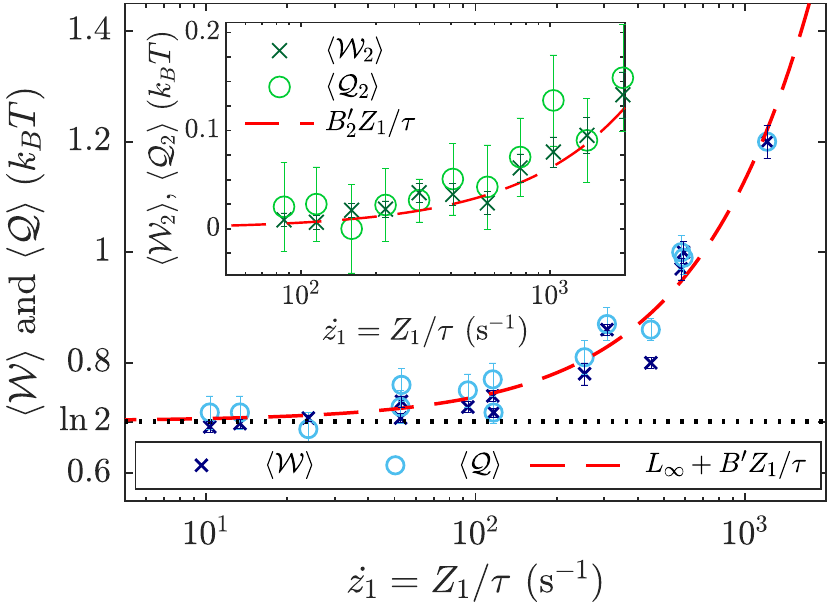}
 \caption{Energy cost from erasure protocols at different speeds $\dot{z}_1=Z_1/\tau$, with $Z_1\sim 5$. Experimental data (blue) are fitted by $L_\infty + B' Z_1 /\tau$ (dashed red), with $B' = (437 \pm 27)\,\SI{}{\micro s}$ and $L_\infty=0.695 \pm 0.012$. (Inset) Experimental mean work and heat during stage 2 (green) at different speeds. The theoretical prediction $B_2' Z_1/\tau$ (dashed red) with $B_2' = Z_1 / Q \omega_0$ works perfectly (no adjustable parameters).}
 \label{Wrampe}
\end{figure}

After providing {experimental evidence that our system matches the LB in the quasi-static limit}, we study how fast the procedure can be performed before paying an extra energetic cost for the erasure. We therefore repeat the experiment for increasing speeds $\dot{z_1}=Z_1/\tau$~\cite{suppmat,note}, and report the results in Fig.~\ref{Wrampe}. The initial distance between the wells may vary slightly from one set of experiments to the other ($Z_1 = 4.5$ to 6, but $Z_1$ is constant for the 2000 experiments of any given $\tau$), so we use the speed $Z_1/\tau$ to represent how far from a quasi-static protocol we stand. As expected, the slow procedures meet the LB (Figs.~\ref{cycle_lent} and \ref{dW_lent} present the slowest one), and quick ones require an extra cost. For finite $\tau$, it has been reported in earlier demonstrations of the LB~\cite{Berut2012,Bech2014} that $\langle \W \rangle \sim \ln 2 + B/\tau$, with $B$ a constant depending on the system and applied protocol. More generally, this $1/\tau$ asymptotical behavior is expected for the mean stochastic work or heat for finite time transformations both in overdamped~\cite{Sekimoto,Aurell_2012,Muratore-Ginanneschi-2017} and underdamped~\cite{Gomez-Marin-2008} systems. This suggests a fit of our results with $L_\infty+B' \dot z_1 = L_\infty + B' Z_1/\tau$. It leads to $B' =(437\pm27)\,\SI{}{\micro s}$ and $L_\infty=0.695 \pm 0.012$ which validates, again, with great accuracy, the Landauer principle. It is noteworthy that using protocols of only $\tau_f \sim \SI{100}{ms}$  (with $\tau= \SI{30}{ms}$ and $\tau_0=3\tau_r=\SI{7.5}{ms}$), the energy cost of erasure is only 10\% larger than the LB. We didn't explore cycles faster than $\tau=6f_0^{-1}$ per ramp, which are already only twice the relaxation time $\tau_r = \frac{1}{\pi}Q f_0^{-1}$.

One may wonder if the extra cost at high speed is due either to the erasure protocol itself, or to the damping losses during the ramps. In the inset of Fig.~\ref{Wrampe}, we plot the contribution $\W_2$ or $\Q_2$ of the ramp of stage 2. In a first approximation (neglecting transients), the cantilever will follow the well center $z_1$ at speed $\dot{z_1}$, while experiencing a viscous drag force $\gamma \sigma \dot{z_1}$. We thus expect the ramp cost to be $\langle \W_2 \rangle = \langle \Q_2 \rangle \sim B_2' Z_1/\tau$, with $B_2' = Z_1 / Q \omega_0$. This approximation with no adjustable parameters is reported in the inset of Fig.~\ref{Wrampe}, and matches perfectly the experimental data. We compute $B_2' \sim \SI{63}{\micro s}$ for $Z_1\sim5$: the ramp contribution is not enough to explain the extra cost of fast erasures, since $2B_2$ (one contribution for each stage) would explain only $\SI{30}{\%}$ of the overhead to $\ln 2$.

As a conclusion, let us summarize the highlights of this work. We demonstrate in this letter the benefits of exploring stochastic thermodynamics with an underdamped micro-oscillator: owing to its fast dynamics and small relaxation time, we can accumulate large statistics and lower uncertainties for the measured quantities. The use of a high-precision interferometer coupled with a simple feedback loop allows for a well-defined and tunable double-well potential energy. Combining these two features, we demonstrate experimentally that the Landauer bound can be reached with a very high accuracy in a short time. Furthermore, we can check every step with analytical predictions. Inertia has no effect on the limit, which is reached for just a few natural relaxation times of the oscillator. Notice that, by using oscillators with a higher resonance frequency, the relaxation time could be considerably reduced. Thus, the present approach is promising to further explore stochastic thermodynamics, with unprecedented statistics and more complex behaviors brought on by the inertia. 

\acknowledgments 
\noindent \textbf{Acknowledgments} This work has been financially supported by the Agence Nationale de la Recherche through grant ANR-18-CE30-0013
and by the FQXi Foundation, Grant No. FQXi-IAF19-05, “Information as a fuel in colloids and superconducting quantum circuits.”

\textbf{Data availability} The data that support the findings of this study are openly available in Zenodo at \url{https://doi.org/10.5281/zenodo.4626559}~\cite{Dago-2021-DatasetPRL}

\bibliographystyle{apsrev4-2}
\bibliography{UDLandauer}

\begin{thebibliography}{40}
\bibitem [{not()}]{note}%
  \BibitemOpen
  \href@noop {} {}\bibinfo {note} {$\langle \W \rangle$ does not depend on
  $\tau_f$ but on $\tau$ only. Indeed as $\dot z_1=0$ outside the ramps period
  then in Eq.~\eqref{WandQ1} the integration limits reduces to
  $2\tau$~\cite{suppmat}.}\BibitemShut {Stop}%
\end{thebibliography}

\end{document}


\title{\textsc{Supplemental material} - Information and thermodynamics: fast and precise approach to Landauer's bound in an underdamped micro-mechanical oscillator}
\author{Salamb\^{o} Dago}
\author{Jorge Pereda}
\author{Nicolas Barros}
\author{Sergio Ciliberto}
\affiliation{Univ Lyon, ENS de Lyon, Univ Claude Bernard Lyon 1, CNRS, Laboratoire de Physique, F-69342 Lyon, France}
\author{ Ludovic Bellon}
\email{ludovic.bellon@ens-lyon.fr}
\affiliation{Univ Lyon, ENS de Lyon, Univ Claude Bernard Lyon 1, CNRS, Laboratoire de Physique, F-69342 Lyon, France}

\maketitle

\section{Feedback and double well potential}

To create the virtual double well potential energy $U(z,z_0,z_1)$, we use a feedback loop as sketched in Fig.~1 of the letter, which must meet two criteria: no hysteresis ($z_0$ is independent of $z$), and negligible delay $\tau_\mathrm{FB}$. The latter necessitates a loop response time much shorter than the natural response time of the system. This is achieved with a high-speed comparator ($\tau_\mathrm{FB} \sim \SI{10}{ns}$) and a slow cantilever (natural free oscillation period $f_0^{-1}\sim \SI{0.8}{ms}$). Implementing a comparator without any hysteresis is less straightforward, since we must prevent the detection noise of the system from causing the comparator to switch back-and-forth when the cantilever is near the threshold. Indeed, this fast switching could result in a mean voltage $V$ actually stabilizing the cantilever around $z_0$ ! We solve this issue by implementing a temporal lock-up that prevents the comparator from switching back for $f_0^{-1}/4$: if the oscillator crosses the threshold, it will evolve in the new well for at least a quarter of its natural period. Consequently, the oscillator has time to reach the well center, far enough from $z_0$ that detection noise cannot cause subsequent switching. By the time the dynamics brings the cantilever back to the threshold, the comparator will be functional again. 

\begin{figure}[t]
 \includegraphics[width=80mm]{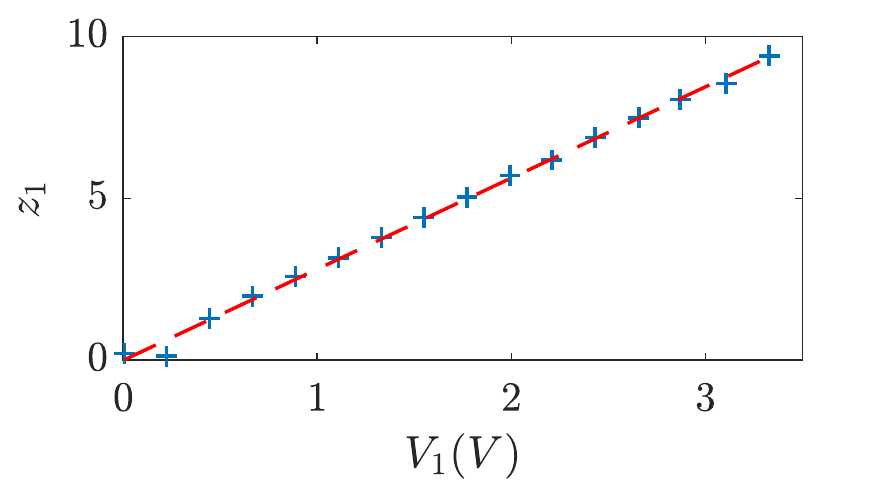}
 \caption{Half distance between the wells $z_1$ returned by the fit of the static probability density function of the cantilever, as a function of the adjustable voltage $V_1$. The linear fit (dashed red) is in good agreement with the experimental points (blue).}
 \label{V1_Z1}
\end{figure}

The distance between the wells $2 z_1$ is driven by the voltage $V_1$. When $V_1\gg V_0$ we can consider that the electrostatic force $F$, and as a consequence $z_1$, is proportional to $V_1$. To confirm this linear behavior, we plot in Fig.~\ref{V1_Z1} the half distance between the wells $z_1$ returned by the fit of the static distribution function of positions $P(z,z_0,z_1)\propto e^{-U(z,z_0,z_1)}$, as a function of the voltage $V_1$. The expected linearity is checked, and $z_1$ can be precisely set during the experiments.

\begin{figure}[b]
\vspace{-11mm}
 \includegraphics{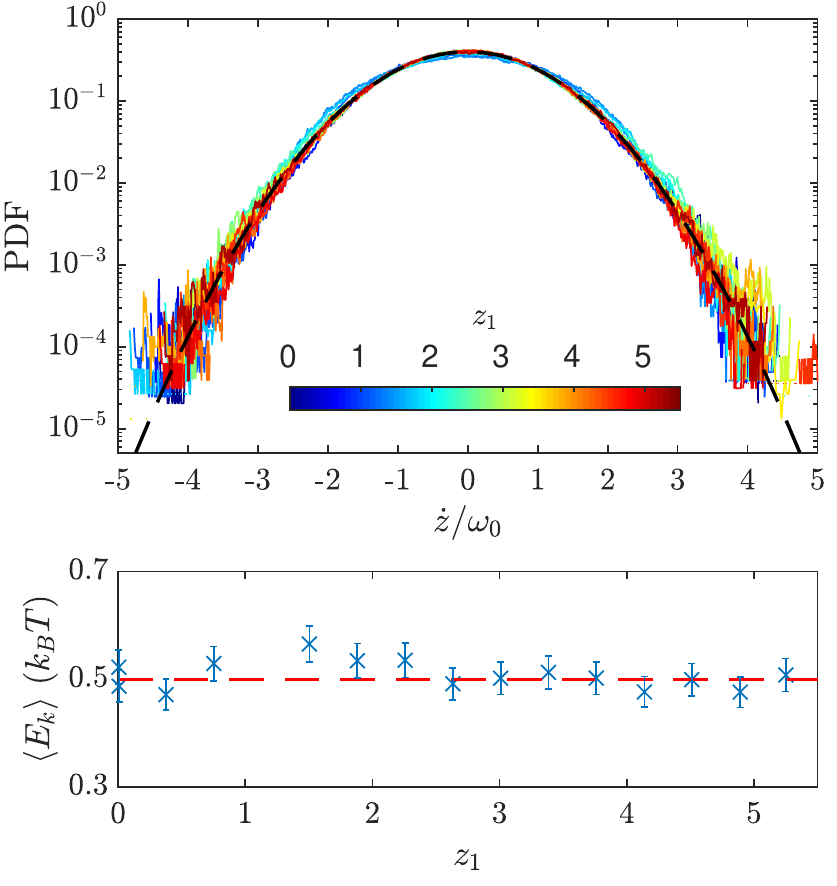}
 \caption{(Top) PDF of the dimensionless velocity of the cantilever $\dot z /\omega_0$ for increasing values of $z_1$ (from $0$ in blue to $5.5$ in red) imposed by the feedback loop. The blue curve corresponds to $z_1=0$ (no feedback). The red line corresponds to very separated double wells: $z_1=5.5$. For each distance between the wells, the experimental PDF fits to the standard normal distribution expected (dashed black). (Bottom) Mean kinetic energy $\langle E_k\rangle = \frac{1}{2}\langle \dot z^2 \rangle/ \omega_0^2$ for increasing distance between the wells. $E_k$ is not influenced by the feedback loop, and it matches the value one would expect from the equipartition theorem.}
 \label{v2_Z1}
\end{figure}

One may wonder whether the feedback loop perturbs the system and impacts its exchange with the thermal bath, and therefore its kinetic energy $E_k = \frac{1}{2} \dot{z}^2/\omega_0^2$. Figure \ref{v2_Z1} aims at showing that the system behaves as in a real potential, whatever the strength of the feedback. A good indicator that this is the case is whether $\langle E_k \rangle$ remains equal to $1/2$ according to the energy equipartition. The experimental results confirm that the velocity distribution remains gaussian and that $\langle E_k\rangle = 1/2$ for any $z_1$.

\section{Stochastic work and heat expressions}
The underdamped system evolves in a potential $U(x,x_i(t))$, where $x_i(t)$ are the external control parameters of the potential. Noting that
\begin{equation}
\dot{U}(x,x_i) = \frac{\partial U}{\partial x}(x,x_i) \dot{x} + \sum_i \frac{\partial U}{\partial x_i}(x,x_i) \dot{x_i},
\end{equation}
we obtain:
\begin{equation}
\dot E_k +\dot U =   \sum_i \frac{\partial U}{\partial x_i} \dot x_i +\frac{\partial U}{\partial x} \dot x + \frac{dE_k}{dt}
\end{equation}
Integrating this equation leads to the energy balance of the system (with the positive exit heat flux convention)
\begin{equation}
\Delta U+\Delta E_k=\W-\Q,
\label{energy}
\end{equation}
where the heat and the work have been defined as (following Refs.~\onlinecite{sek10,Sekimoto}):
\begin{align}
W&=\int \sum_i \frac{\partial U}{\partial x_i} \dot x_i \dd t \\
Q&=- \int  \frac{\partial U}{\partial x} \dot x \dd t - \Delta E_k
 \label{WandQ0}
\end{align}
In our experiment, the potential energy $U$ is given by
\begin{align}
U(z,z_0,z_1)=\frac{1}{2} \big(z-S(z-z_0)z_1\big)^2.
\label{eq_U(z,z0,z1)}
\end{align}
The contribution of $(\partial U/\partial z_0) \dot z_0 \propto \delta (z-z_0)z_1\dot z_0$ disappears in the work expression: in our protocol $\dot z_0 =0$, except at two particular times corresponding to the beginning and end of stage 2. The former corresponds to $z_1=0$, and the latter corresponds to the situation where $z$ and $z_0$ cannot cross, which constrains this term to 0. In the heat expression, the contribution of the $\partial S/\partial z$ term in $(\partial U / \partial z) \dot z $ is proportional to $\delta (z-z_0) z_1 z \dot z $ and vanishing as well: assuming that $z(t)$ and $z_0(t)$ intersect at $t=t_0$, this contribution is
\begin{align}
\int_{\sim t_0} \delta(z-z_0) z_1 z \dot z \dd t = \frac{z_1(t_0) z(t_0)\dot z(t_0)}{\dot z(t_0) -\dot z_0 (t_0)} = 0,
\label{term}
\end{align}
since during our protocol $z$ and $z_0$ only intersect during stage 1 where $\dot z_0=0$ at all times.

Eventually $\W$ and $\Q$ write:
\begin{align}
\W& = - \int_0^{\tau_f} \big(S(z-z_0)z-z_1\big) \dot z_1 \dd t \label{WandQ1}\\
\Q& = \int_0^{\tau_f} \big(S(z-z_0) z_1-z \big) \dot z \dd t - \frac{1}{2\omega_0^2} \Big[ \dot{z}^2 \Big]_0^{\tau_f}. \label{WandQ}
\end{align}
It should be noted that in the computation of the mean dissipated heat, we did not include the kinetic term which vanishes on average. These general expressions can be used to deduce work and heat during stage 1 or during stage 2, by adapting the integration bounds. It is straightforward to compute $\W$, since $\dot z_1=0$ outside the ramps. For $\Q$, we add at least $\SI{5}{ms}$ after reaching the final state, i.e.~at least 2 relaxation times so that the system is very close to equilibrium.

\vspace{5mm}
\textbf{During stage 1} we have $z_0=0$ and $z_1=Z_1(1-t/\tau)$, the potential energy is therefore simply
\begin{equation}
U_1(z,z_1)= \frac{1}{2} \big(|z|-z_1)^2.
\end{equation}
The work is then expressed:
\begin{align}
\W_1 & = - \int_0^{\tau} \big(|z|-z_1\big) \dot z_1 \dd t \\
& = \int_0^{Z_1} (\lvert z\rvert - z_1) \dd z_1.
\end{align}
For slow protocols ($\tau f_0 \gg Q$), we can assume that the PDF $P(z)$ satisfies at all times the static expression given by the Boltzmann distribution:
\begin{align}
P(z) = \frac{1}{\Z} e^{-U_1(z,z_1)} = \frac{1}{\Z} e^{-\frac{1}{2}(|z|-z_1)^2} \label{Px},
\end{align}
where
\begin{equation}
\mathcal{Z}=\sqrt{2\pi}\left(1+\erf(z_1/\sqrt{2})\right).
\end{equation}
$P(z)$ is even, so we can deduce that: 
\begin{align}
\langle \lvert z \rvert \rangle &= \int_{-\infty}^\infty \lvert z \rvert P(z) \dd z = 2 \int_{0}^\infty z P(z) \dd z \\
&=\frac{2}{\Z} \int_{0}^\infty z e^{-\frac{1}{2}(z-z_1)^2}\dd z \\
&=\frac{2}{\Z} \left(z_1\int_0^\infty P(z) \dd z - \Big[e^{-\frac{1}{2}(z-z_1)^2}\Big]_0^\infty\right) \\
&=z_1+\sqrt{\frac{2}{\pi}} \frac{e^{-\frac{1}{2}z_1^2}}{1+\erf (z_1/\sqrt{2})}
\end{align}
All in all, we have the mean work in stage 1: 
\begin{align}
\langle \W_1 \rangle &=\langle \int_0^{Z_1} (\lvert z\rvert - z_1) \dd z_1 \rangle \\
&=\sqrt{\frac{2}{\pi}} \int_0^{Z_1}  \frac{e^{-\frac{1}{2}z_1^2}}{1+\erf (z_1/\sqrt{2})} \dd z_1\\ \label{W1power}
&=\ln(1+\erf(Z_1/\sqrt{2})) \underset{Z_1 \gg 1}{\longrightarrow} \ln 2.
\end{align}
For a robust 1-bit of information, and equivalently, for a complete and reliable erasure, we should take $Z_1\gg 1$ in the initial state. As a result, for quasi-static erasure processes of 1 bit, $\langle \W_1 \rangle =k_B T \ln 2$.

During the quasi-static merging of the wells, we can also compute the average instantaneous power:
\begin{align}
\langle \frac{\partial U}{\partial z_1} \dot z_1 \rangle & = - \langle \big(|z|-z_1\big) \dot z_1\rangle \\
& = \sqrt{\frac{2}{\pi}} \frac{e^{-\frac{1}{2}z_1^2}}{1+\erf (z_1/\sqrt{2})} \frac{Z_1}{\tau}.
\end{align}
This expression is plotted in Fig.~4 of the letter, and it closely follows the experimental data.

\vspace{5mm}

\textbf{During stage 2}, the cantilever no longer sees the second well and $S(z-z_0)=-1$, so that the potential simplifies into: 
 \begin{align}
 U_2(z,z_1)&=\frac{1}{2}(z+z_1)^2.
\end{align}
The mean work becomes 
\begin{align}
\langle \W_2 \rangle &=\langle - \int_0^{\tau} \big(z+z_1\big) \dot z_1 \dd t \rangle \\
&= - \int_0^{\tau} \big(\langle z \rangle +z_1 \big) \dot z_1 \dd t.
\end{align}
During the quasi-static translation, $\langle z \rangle=-z_1$, and so $\langle \W_2 \rangle=0$.

\section{Fast erasure protocols}

\begin{figure}[b]
 \includegraphics[width=85mm]{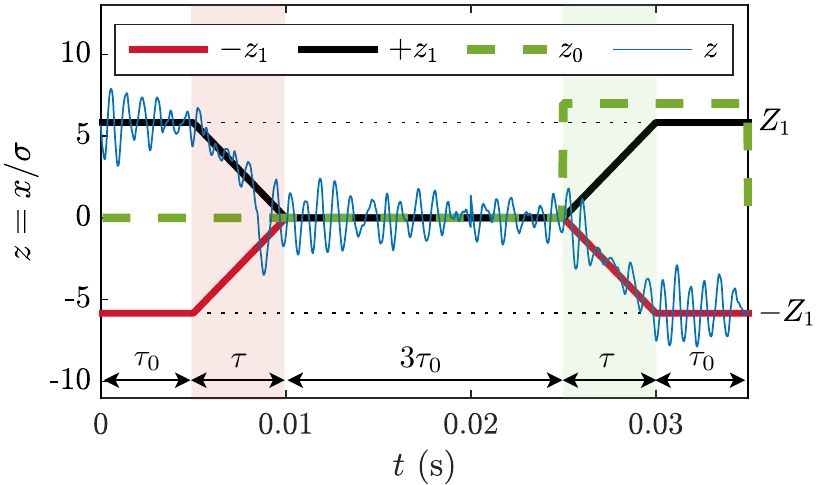}
 \caption{Fast erasure protocol: the stage 1 and stage 2 duration is $\tau=\SI{5}{ms} $, that is to say $f_0\tau =6.4$.}
 \label{cycle_rapide}
\end{figure}

\begin{figure}[h]
 \includegraphics[width=80mm]{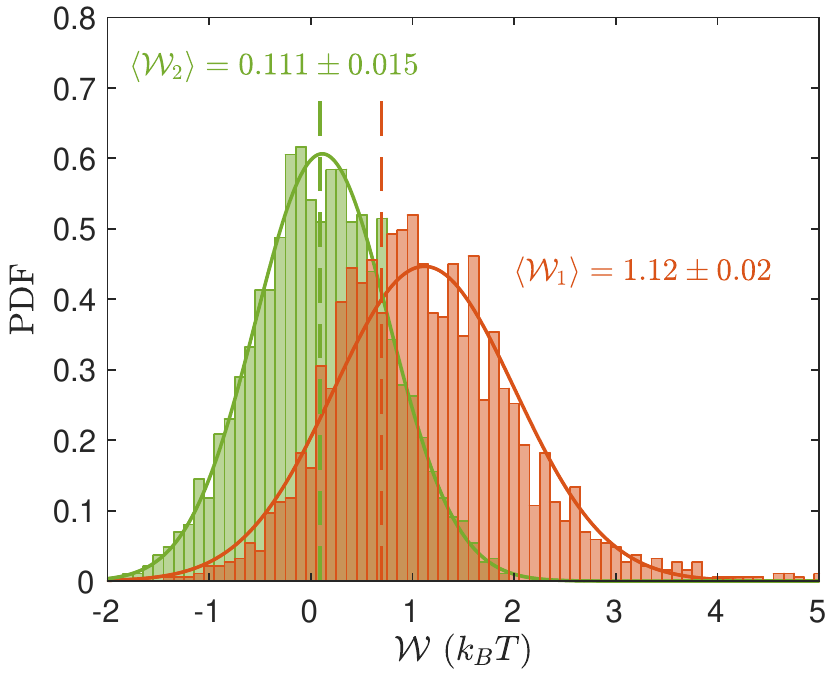}
 \caption{Work distribution of $2000$ iterations of the fast protocol of Fig.~\ref{cycle_rapide}. For fast erasure, stage 2 contribution is no longer negligible: $\langle \W_2 \rangle = 0.111\pm0.015$ is close to the expected asymptotic behavior $Z_1^2/(Q \omega_0 \tau)=0.091$, represented by the dashed green line. Stage 1 mean work $\langle \W_1 \rangle = 1.12\pm 0.02$ is now clearly above the LB represented by the red dashed line.}
 \label{W_histo_rapides}
\end{figure}

\begin{figure}[h]
 \includegraphics[width=85mm]{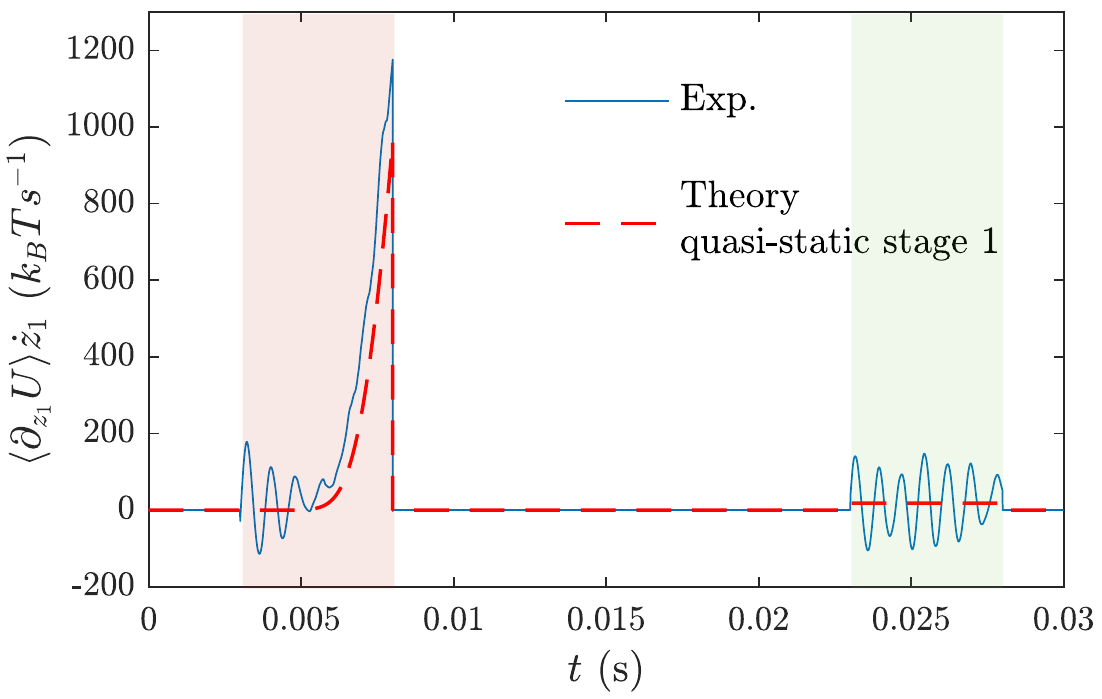}
 \caption{Time evolution of the mean power over $2000$ trajectories following the fast protocol of Fig.~\ref{cycle_rapide}. The red dashed line corresponds to the quasi-static regime prediction for stage 1 (Eq.~\ref{W1power}) and to the asymptotic mean power value for stage 2:   $Z_1^2/(Q \omega_0 \tau^2)$.}
 \label{dWrapide}
\end{figure}

In this section, we reproduce the results presented in Fig.~3 and Fig.~4 of the article, but for protocols that can no longer be considered as quasi-static. For fast protocols corresponding to a few oscillations of the cantilever such as the one in Fig.~\ref{cycle_rapide}, the system can switch only once or twice between the wells. The work required is above the Landauer limit as pointed out in Fig.~\ref{W_histo_rapides}. Fig.~\ref{dWrapide} details the mean power required during the process: stage 1 power exceeds the quasi-static curve and the stage 2 contribution no longer vanishes.

\section{Heat and Work distribution}

\begin{figure}
 \includegraphics[width=90mm]{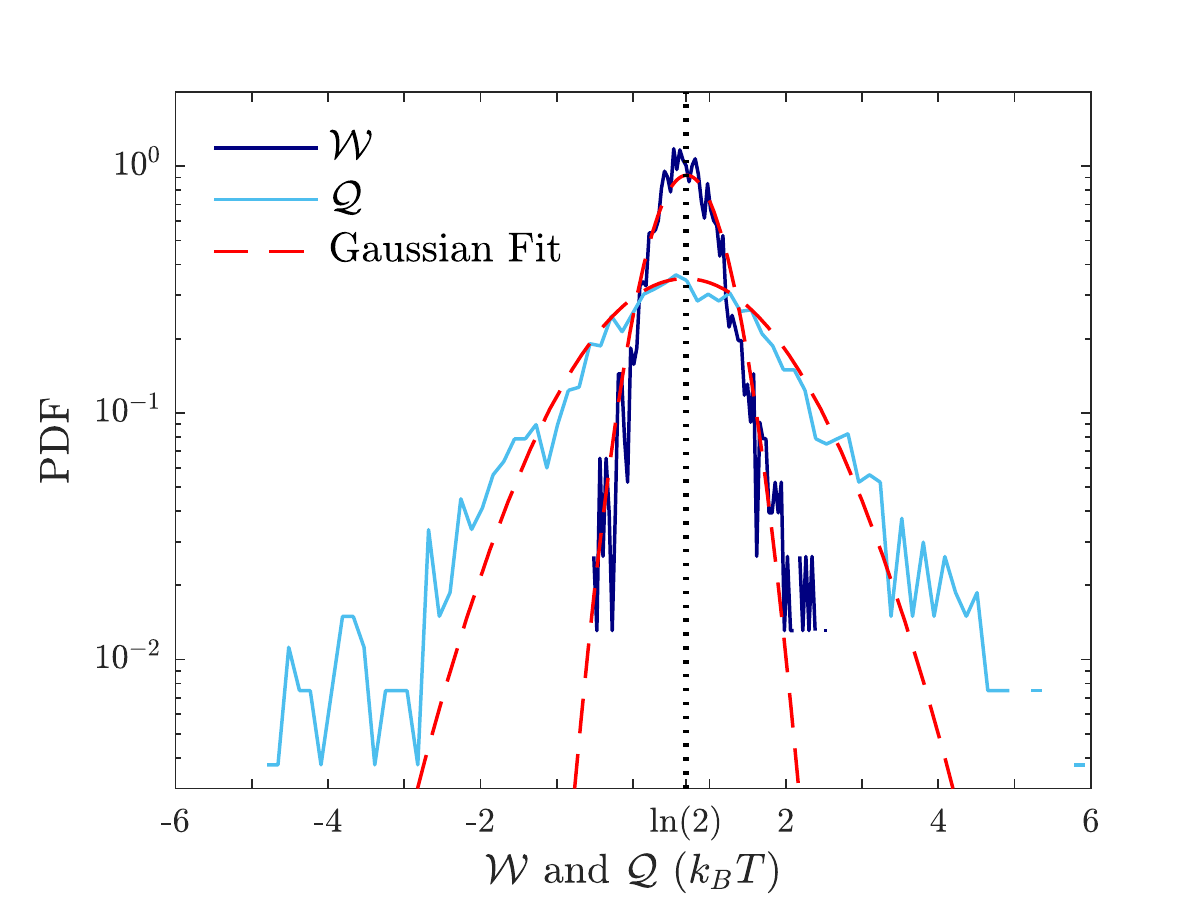}
 \caption{Work (dark blue) and heat (light blue) distributions of $2000$ erasure procedures of Fig.~3 of the letter. The gaussian shape of the work PDF is highlighted by the best fit to a gaussian distribution (dashed red line). The heat is far more dispersed than the work and presents exponential tails. Since it is a slow protocol, both are centered on the LB (dotted black vertical line).}
 \label{W_Q_lent}
\end{figure}

We present in Fig.~\ref{W_Q_lent} the distribution of the work and the heat of $2000$ iterations of the erasure with the slow protocol detailed in the letter Fig.~3. The work distribution is gaussian. Meanwhile, the heat distribution presents exponential tails. This observation is in agreement with the study of Ref.~\onlinecite{Ciliberto_PRX}: $\Q$ is the convolution product of a gaussian and an exponential distribution. Besides, both distributions are centered on the LB because the protocol under study is close to the quasi-static regime.

\bibliographystyle{apsrev4-2}
\bibliography{UDLandauer}